# CONNECTING QUANTUM AND COSMIC SCALES BY A DECREASING LIGHT SPEED MODEL

Juan Casado

*Matgas Research Centre, Universidad Autonoma de Barcelona, Pl. Cívica, P12, 08193, Bellaterra (Barcelona) Spain*

In this paper we point out that the radius $R$, the age $t$ and the mass $M$ of the observable Universe are related to Planck units for length $l_p$, time $t_p$ and mass $m_p$ as

$$\frac{R}{l_p} \approx \frac{t}{t_p} \approx \frac{M}{m_p} \approx 10^{61}.$$

This equality is related, by a cosmological model linking the universal expansion to the speed of light, with a time variation of $c$ and of the Planck's constant as $c \propto \hbar^{-1} \propto t^{-1/3}$, which implies that $l_p$ and $t_p$ increase with time while $m_p$ remains constant. We discuss some of the implications these relationships could have on quantum cosmology and obtain that the Universe entropy associated to the event horizon is constant along its history. Quantum effects were smaller in the past, in such a way that their conflict with relativity about the Planck era vanishes.

## I. INTRODUCTION

Several cosmological models assuming a variable speed of light in vacuo along cosmic history have been proposed and are currently under careful scrutiny (see [1] for a review). In fact, Lord Kelvin already proposed the idea of a variation of $c$ as function of the age of the Universe in 1874. On the other hand, the possible time variation of $\hbar$ has been much less explored. Here, we outline an alternative cosmological model in which the variation of $c(t)$ implies also a change of $\hbar(t)$ in a precise way in order to make the model self-consistent. To undertake this task, we analyse some numerical relations between the cosmic scales set by the radius, $R$, the mass, $M$, and the age, $t$, of the observable Universe, and the quantum scales given by natural Planck units for length, time and mass. These are defined as combinations of $\hbar$, $c$ and the gravitational constant $G$, namely

$$l_p = (G\hbar/c^3)^{1/2} \qquad (1)$$

$$t_p = (G\hbar/c^5)^{1/2} \qquad (2)$$

$$m_p = (\hbar c/G)^{1/2}. \tag{3}$$

Since in these definitions quantum ($\hbar$), electromagnetic ($c$) and gravitational ($G$) features appear, these scales are interpreted as the ones where a theory unifying quantum physics and general relativity (GR) would be indispensable for a description of the underlying processes, thus setting limits on our current ability to describe the early Universe at ages shorter than Planck's time.

Note that if $c$, $\hbar$, or $G$ are allowed to change with time, Planck's units could also vary with time. Here we will analyse the hypothesis that these natural units may change in such a way that the following order-of-magnitude relationships

$$\frac{R}{l_p} \approx \frac{t}{t_p} \approx \frac{M}{m_p} \approx 10^{61}, \tag{4}$$

which are valid for the current values $R_o$, $t_o$ and $M_o$ of our Universe, would be also valid at any time along the cosmic history (through this paper, subscript $_o$ refers to present-day values of the respective parameters).

In section II, we justify our claims for this hypothesis and we set it in a theoretical and historical context. In section III, it is shown that the hypothesis follows quite directly if $c$ decreases and $\hbar$ increases with time in such a way that $\hbar c$ remains constant. We also study the thermodynamic implications of our hypothesis, which involve the constancy of the entropy of the observable Universe along its expansion. Finally, in sections IV and V we examine some other theoretical explorations consistent with the proposals of the present work, and some observational clues pointing to their plausibility.

## II. THE COVARIANCE HYPOTHESIS

First of all, we calculate the ratio between the current radius of the observable Universe ($R_o$) and $l_p$, between the age of the Universe ($t_o$) and $t_p$, and between the mass of the observable Universe ($M_o$) and $m_p$. We carry out such calculations in the context of two different cosmological models: first, one involving decreasing light speed, and, second, a standard model for a flat Universe. In both of them the order-of-magnitude of the results is the same.

In a previous work [2] a non-standard cosmological model with Decreasing Light Speed (shortened as *DeLightS*) was presented. The main relationships obtained between $R$, $M$, and $t$ were

$$R = \frac{3}{2}ct \tag{5}$$

$$2GM = Rc^2, \tag{6}$$

where $R$ and $c$ are functions of $t$. Note that the factor 3/2 in equation (5), which is absent in the standard model, comes from the way $c$ changes with time in *DeLightS* model, namely as $c \propto t^{-1/3}$. Values of $R_o = 2.0 \ 10^{26}$ m and $t_o = 1.4 \ 10^{10}$ years were deduced from a revised redshift-distance relationship [2]. From $R_o$ and equation (6) a mass $M = 1.4 \ 10^{53}$ kg is obtained. According to *DeLightS* model, our event horizon and the space-time frame expand at the same rate, given by $c(t)$; i.e. no particle nor wave can cross over the cited horizon. As a consequence, $M$ remains constant, independent of $t$ and $R$, and finite, even in the case of an infinite Universe. Note that, in this model, Hubble parameter is given by $H = c/R = 2/3t$.

Dividing the actual value $R_o$ quoted above by the Planck length $l_p$ one obtains

$$R_o/l_p = 1.2 \ 10^{61}. \tag{7}$$

This huge number can be interpreted as the scale factor of universal expansion since the Planck era. Let us do the same with times $t_o$ and $t_p$

$$t_o/t_p = 0.8 \ 10^{61}. \tag{8}$$

Finally, from masses $M$ and $m_p$ we obtain

$$M/m_p = 0.6 \ 10^{61}. \tag{9}$$

Note that these three dimensionless ratios are of the same order, roughly $10^{61}$. The minor differences among them can be ascribed to numerical factors of order unity, as discussed below.

The above coincident results are not mere artefacts from *DeLightS* model. In order to prove it, let us consider for instance one of the most favoured cosmologies nowadays: a standard model for a flat universe with $H_o = 70$ km s$^{-1}$ Mpc$^{-1}$ = 2.3 $10^{-18}$ s$^{-1}$. The Hubble age would be $t_o = 1/H_o = 1.4 \ 10^{10}$ years, so that $t_o/t_p = 0.8 \ 10^{61}$. The present-day radius would be $R_o = ct_o = 1.3 \ 10^{26}$m, so that $R_o/l_p = 0.8 \ 10^{61}$. Finally, the mass density of such Universe can be obtained as $\rho_o = \rho_c = 3H_o^2/8\pi G \cong 10^{-26}$ Kg m$^{-3}$, which leads to a mass within the observable Universe $M_o = 4\pi R_o^3 \rho_o/3 \cong 9 \ 10^{52}$ Kg, and therefore $M/m_p \cong 0.4 \ 10^{61}$. Although these ratios are somewhat lower, on average, than those obtained in (7), (8) and (9), they still have the same order of magnitude, so that the numerical estimation (4) also holds for standard model solutions.

In this paper, we propose that the order-of-magnitude agreement denoted by the approximation (4) holds not only at present time, but also for any cosmic age, and contains some significant information about the physical properties of our Universe. We will refer to this space-time-mass relationship merely as the *covariance hypothesis*. Of course, this statement would be not viable in models considering that $c$, $\hbar$ and $G$ are constant, because in that case the Planck units would be also constant, whereas $R$ and $t$ increase with time, thus making untenable that the ratios $R/l_p$ and $t/t_p$ are constant. However, as we will show, our hypothesis is consistent in a model with $c$ and $\hbar$ varying with time in a definite way.

Similar coincidences in other quantities were a matter of surprise already eight decades ago, when Eddington noted that two large dimensionless numbers that characterise our Universe are approximately equal, namely

$$\alpha' = \frac{e^2}{Gm_e m_n} \approx 10^{40} \tag{10}$$

$$\beta = \frac{ct}{e^2/m_e c^2} \approx 10^{40}, \tag{11}$$

where $e$ and $m_e$ are the electron charge and mass, and $m_n$ refers to the mass of a nucleon. The first of these numbers is the ratio between the electromagnetic force by which a proton attracts an electron and their respective gravitational attraction force, whereas the second number is the ratio between the size of the observable Universe and the classical radius of an electron. This coincidence was regarded by most researchers as fortuitous, while a few others, leaded by Dirac, attributed a relevant paper to it, up to the point of proposing a cosmological theory based on this so-called large number hypothesis and postulating a varying $G \propto t^{-1}$ [3]. However, careful observations of planetary and stellar orbits, stellar evolution and cosmic nucleosynthesis show that $G$ has not changed more that 1% along the Universe history, a change much smaller than Dirac's prediction [4]. By the way, let us note that if Dirac had considered the possibility of a varying $c(t)$, assuming the remaining quantities in equation (11) to be actually constant, he would have obtained immediately the result $c \propto t^{-1/3}$, even in case that the big numbers coincidence was merely fortuitous. This dependence is in agreement with *DeLightS* model, which gives a precise physical meaning to it.

Let us algebraically derive the expression (4) from *DeLightS* model in order to prove the validity of (4) for any cosmic age in its framework. From definition (1) and taking $R$ from (6) it is straightforward to obtain

$$\frac{R}{l_p} = \frac{2MG^{1/2}}{\hbar^{1/2}c^{1/2}} = \frac{2M}{m_p}. \tag{12}$$

In a similar way, combination of (2), (5) and (6) yields

$$\frac{t}{t_p} = \frac{2Rc^{3/2}}{3G^{1/2}\hbar^{1/2}} = \frac{4MG^{1/2}}{3\hbar^{1/2}c^{1/2}} = \frac{4M}{3m_p}. \tag{13}$$

As can be seen, $R/l_p$ and $t/t_p$ are reduced to (9), except for the numerical factors derived from equations (5) and (6). The equalities between the former ratios do not imply by themselves that these ratios are constant. We propose that this constancy is due to the fact that $M$ is constant (as mentioned above) and that $G$ and $\hbar c$ are unvarying (as shown below), thus implying that $m_p$ itself is constant. Thus, it follows that, for any time,

$$\frac{R}{2l_p} = \frac{3t}{4t_p} = \frac{M}{m_p} \cong 0.6 \times 10^{61}. \tag{14}$$

The *covariance hypothesis* thus states that equations (4) and (14) are not only valid in our epoch, but considers instead that the Universe always has roughly the same radius, age and mass when measured in Planck units, i.e. that the dimensionless number 0.6 $10^{61}$ has a fundamental physical meaning. Note that the simple formula (4) seems to nicely connect the microcosmos with the cosmos, linking the smallest physical units of length and time with the largest dimensions of space and time of the whole observable Universe. Moreover, it suggests a direct connection between quantum physics, related to Planck units, and cosmology. At present, this cannot be shown a priori from any previous theory, but its formal elegance invites to pay some attention to its possible consequences. In the following, we explore the implications of this hypothesis.

Let us note that the constancy of $R/l_p$ implied by (4) has a thermodynamic interpretation in terms of the constancy of the entropy of the Universe. Indeed, the Bekenstein [5] and Hawking [6] black hole entropy is proportional to the area of its event horizon (i.e. to its squared radius) relative to the squared Plank length. Applying the Bekenstein-Hawking hypothesis to the cosmic horizon, one would have, in units of Boltzmann constant $k$:

$$\frac{S}{k} \approx \left(\frac{R}{l_p}\right)^2 \approx 10^{122}. \tag{15}$$

This is in fact the order of magnitude of the Universe entropy as estimated by Lloyd [7] on different grounds. The constancy of $R/l_p$ would thus be equivalent to a reversible adiabatic expansion of the Universe. The fact that an equation originally derived for black holes can also be applied to the overall Universe [8] further supports the applicability of equation (6) to both scenarios. Therefore, the Universe might behave as a huge black hole, at least in the sense that its entropy corresponds to the horizon of the observable Universe, a system also bound by gravity. This idea fits well with *DeLightS* model, were neither particle nor wave can cross that horizon and it may find yet another basis on the holographic principle [9-12]. Analogously, the entropy of any black hole of constant mass $m_H$ would also remain constant, because its Schwarzschild radius ($r_H = 2Gm_H/c^2$) increases at the same rate as $l_p$ ($l_p = Gm_p/c^2$) increases. Note that the *covariance hypothesis* implies an expansion of all black holes by the mere fact that $c$ is decreasing as time goes by.

On the other hand, the entropy due to radiation or, more precisely, due to all the particles present in the Universe is much smaller than the entropy ascribed to the cosmic horizon, and behaves as $S_\gamma = S^{3/4} \sim 10^{91} \ll 10^{122}$ [7,12]. This entropy 'gap' between $S_\gamma$ and $S$ has been attributed to the gravitational field contribution.

Let us finally mention some surprising 'coincidences' such as the matter and dark energy dominance today [13], the closeness of the present radius of the Universe ($R_o$) to the distance travelled by light during $t_o$ [2], or the similarity of the empirical density of matter today, the critical density of the Universe and the density of a black hole of radius $R_o$ [3,14,15]. In our model, these coincidences are interpreted not as mere chances, but as clues to the nature of the Universe. For instance, the last mentioned density coincidence is a prediction of *DeLightS* for any cosmic time due to the formal agreement of equation (6) with the Schwarzschild solution for GR describing the radius of the event horizon of a spherical black hole [2].

### III. A SLOWLY INCREASING PLANCK'S CONSTANT

The constancy of the ratios appearing in equations (12) and (13) implies that, assuming that $G$ and $M$ are actually constant, the product $\hbar c$ should be also constant; since in *DeLightS* model $c$ is decreasing with time, it follows that $\hbar$ should correspondingly increase as $c^{-1}$. An argument to understand the consistency of this change in $\hbar$ is obtained by considering the De Broglie equation for a photon $\lambda = \hbar/p$.

The average energy of the photons filling the Universe (mainly cosmic background radiation) is proportional to its temperature, which in turn is proportional to $R^{-1}$. From equation (6), we get immediately $R^{-1} \propto c^2$. This implies for the photon energy a dependence $E \propto c^2$, a result also obtained in other variable speed of light (VSL) model [16]. Therefore, for the photon momentum we obtain:

$$p = \frac{E}{c} \propto c \propto R^{-1/2} \ . \tag{16}$$

On the other hand, the photon's wavelength stretches as the scale factor of the Universe: $\lambda \propto R$. Therefore

$$\hbar = p\,\lambda \propto R^{-1/2}\,R = R^{1/2} \ . \tag{17}$$

This dependence is exactly the reciprocal to that of $c$, so that the product $\hbar c$ ($\propto R^{1/2} R^{-1/2}$), known as the conversion constant, seems to be really independent of the Universe scale, either if $\hbar$ and $c$ were constant (standard theory) or if both were changing in such a way that their coupled variations cancel (*DeLightS* model). This result may be seen as a consequence of the *covariance hypothesis* and, thus, a part of it.

One of the consequences of the constancy of $\hbar c$ is that the fine structure constant $\alpha$ should be indeed a universal constant, provided the electron charge $e$ does not vary with time. If the reports of Webb *et al*. [17] on minute changes in $\alpha$, of the order of $10^{-16}$ parts per year, were independently confirmed, they might be indicating a small change in $e$ [18] instead of a change in $c$ [19], which would be much smaller than predicted by *DeLightS*. Anyway, a variation of $e$ would not affect the *covariance hypothesis*, since $e$ does not appear either in equation (4) or in Planck units.

Barrow has recently pointed out some problems of 'naïve' VSL models [20]. For instance, the quantum wavelength of massive particle state of mass $m$, defined as $\lambda = \hbar/mc$, could grow to exceed the scale of the particle horizon, given by $r = ct$ and would evolve to become acausal separate universes (!). In contrast, in *DeLightS* model, with $\hbar$ varying as $c^{-1} \propto t^{1/3}$, $\lambda$ will grow as $t^{2/3}$, i.e. at the same rate as $r$, in such a way that the quantum wavelength will be always smaller than the particle horizon, either in the future or in the past. This would not happen if $\hbar$ is constant, because in this case the quantum wavelength would scale as $t^{1/3}$ and the horizon radius as $t^{2/3}$, so that for t→0, the quantum wavelength would become higher than the particle horizon (!). A similar puzzle arises when considering the growth of primordial black holes in previous VSL models, i.e. the black hole horizon grows faster than the particle horizon [20]. *DeLightS* avoids this problem as well since the black hole radius and the horizon radius have exactly the same dynamics, given by equation (6), so that any black hole of mass smaller than $M$ will never reach the event horizon radius $R$.

According to (3), the constancy of $\hbar c$ and $G$ is equivalent to the constancy of $m_p$. This feature is especially satisfactory within *DeLightS* and some other models featuring finite universes, where $M$ is also considered constant. On those grounds, we said in (14) that the ratio $M/m_p$ has to be a universal constant. Let us explore now the consequences for $l_p$ and $t_p$. If $M/m_p$ is constant, equations (4) and (14) lead to the conclusion that the dimensionless ratios (7) and (8) do not depend on the age or the size of the Universe. Therefore, $l_p$ and $t_p$ are both varying proportionally to $R$ and $t$, respectively, as the Universe evolves. This statement may seem surprising, but the evolution rates of $\hbar$ and $c$, derived above, have exactly the values required to make these ratios constant for any cosmic age. To be sure, let us first consider $l_p$. From (1), (17) and taking into account that $c \propto R^{-1/2}$ it is immediate that

$$l_p^2 = \frac{G\hbar}{c^3} \propto \frac{R^{1/2}}{R^{-3/2}} = R^2,$$

(18)

so that $l_p \propto R$. In a similar way, for $t_p$ we have

$$t_p^2 = \frac{G\hbar}{c^5} \propto \frac{R^{1/2}}{R^{-5/2}} = R^3,$$

(19)

and taking into account that in *DeLightS* $R^3 \propto t^2$ [2] (as in any flat-universe model with $M$ and $\Omega_M$ constant or in any standard model along the matter-dominated era), we immediately get $t_p \propto t$, as it was anticipated.

## IV. ADDITIONAL CONSIDERATIONS

In this section we will look for further independent support to the *covariance hypothesis*, starting not from its direct consequences, which have been just analysed, but focusing on two other related considerations.

## A. An equation by Teller

Five decades ago, Teller, looking for a fundamental connection between different 'constants' of nature including $\alpha$, $c$, $\hbar$, $H_o$, the Einstein constant ($\kappa = 8\pi G/c^4$) and the Planck units, found a remarkable equation that in modern notation reads [21]:

$$\kappa \hbar H_o / l_p = 8\pi t_p H_o = \kappa m_p H_o c = \exp(-1/\alpha). \qquad (20)$$

We will not consider the $\alpha$ term in this work because it is not necessary for our purposes. For the first three terms, substituting $\kappa$ and dividing by $4\pi$ we get

$$\frac{2G\hbar H_0}{c^4 l_p} = 2 t_p H_0 = \frac{2G m_p H_0}{c^3} = 1.6 \times 10^{-61}, \qquad (21)$$

as calculated by using $H_o = 1.5\ 10^{-18}\ s^{-1}$ [2, 22].

Due to the numerical value obtained, it is tempting to equalise term by term the inverse of equation (21), generalised to any time by removing the subscript $_0$, with equation (14) proposed in section 2. Concerning length we then have:

$$\frac{R}{2 l_p} = \frac{c^4 l_p}{2G\hbar H}, \qquad (22)$$

and thus, considering (1), we get

$$RG\hbar H = c^4 l_p^2 = cG\hbar, \qquad (23)$$

and therefore we are led to the relation $H = c/R$, which is precisely the first postulate of *DeLightS* [2]. For time we have

$$\frac{3t}{4 t_p} = \frac{1}{2 t_p H}, \qquad (24)$$

which implies $H = 2/3t$, one of the main results of *DeLightS*, also in agreement with Einstein-de Sitter cosmology [22]. Finally, for mass, it is obtained

$$\frac{M}{m_p} = \frac{c^3}{2G m_p H}, \qquad (25)$$

which, considering that $H = c/R$, recovers result (6).

Therefore we conclude that the *covariance hypothesis*, Teller's equation and *DeLightS* model are mutually consistent, which adds likelihood to their correctness. It must be noted, however, that Teller neither extended (20) to any cosmic time, nor considered variations in *c* or in Planck units.

**B. A huge dimensionless constant**

A second consideration is based on the 'coincidences' of other numerical dimensionless combinations of the quantities *R, t, M, G, c* and *ħ,* powered to natural exponents 1, 2 or 3. A convenient way to get these numbers is by squaring or multiplying in pairs the terms of equation (14). For instance, squaring the first term and taking into account the definition (1) we immediately get

$$\frac{R^2 c^3}{4G\hbar} \cong 4 \times 10^{121}. \qquad (26)$$

In a similar way, when multiplying the second by the third term of (14), one obtains

$$\frac{3Mc^2 t}{4\hbar} \cong 4 \times 10^{121}. \qquad (27)$$

Finally, multiplying the third term by itself yields

$$\frac{GM^2}{\hbar c} \cong 4 \times 10^{121}. \qquad (28)$$

Considering the way we have reached them, it is not surprising that all these dimensionless numbers have the same value. Equivalent expressions that yield the same number are $McR/2\hbar$ and $MR^2/3\hbar t$.

It is tempting to attribute to this unique number, perhaps the largest dimensionless constant with a physical meaning, a relevant significance in cosmology. For instance, it 'coincides' with the maximum number of elementary quantum logic operations that the Universe can have performed, as calculated by Lloyd [7] on different grounds (from the Margolus-Levitin theorem). It also agrees with the universal entropy calculated using the Bekenstein-Hawking formalism and the holographic principle, as we have already noted at the end of section 2. The new information the *covariance hypothesis* implies is that both quantities should be actually invariant: neither the information capacity of the Universe, nor its entropy have changed with time, nor will vary in the future. They would be fixed since the number of universal quanta and the total number of particles are constant, despite cosmological age and size keep growing.

In that scenario the second principle of thermodynamics is fulfilled, in contrast to other models, including VSL ones [20]. One cannot exclude, furthermore, a possible anthropic interpretation of this number, which we will explore in the future.

## V. OBSERVATIONAL CONSTRAINTS

The present section is devoted to discuss the compatibility of the *covariance hypothesis* with current observations. Results recently reported by two different teams [23,24] reveal the lack of observable quantum structure of space-time when observing distant objects. Quantum gravity models [6,23,25-28] normally predict that the cumulative effect of Planck-scale phenomenology would produce the loss of the phase of radiation emitted at large distances from the observer. A testable consequence of such models is that the Airy rings of very distant objects should be blurred by the loss of coherence of light travelling through a quantum-structured space-time. However, these predicted effects have not been found at all. For instance, the diffraction patterns from the *Hubble Space Telescope* observations of nearby stars are as sharp as those of SN 1994D, 14 Mpc away. The appearance of a Hubble Deep Field galaxy at $z = 5.34$ and the detection of Airy rings from the active galaxy PKS 1413_135, located at a distance of 1.2 Gpc, show the same behaviour: the observed images are sharp at all scales.

The new observations cast doubt on the physical significance of a Planck length and a Planck time, and have been interpreted as a lack of quantum structure of space and time; i.e. space-time would be perfectly continuous. Some rebuttals of this interpretation have also appeared [29,30]. Another possibility is that time and space vary together at the Planck scale, keeping the phase coherence of light waves as originated [23]. Finally one cannot rule out the eventuality that the cited quantum gravity models are flawed. Recent results on the strong polarisation of gamma rays from GRB 021206 also seem to constrain quantum gravity models in a similar way [31].

Whatever the correct view would be, we want to point out that the above observations can be explained in view of the *covariance hypothesis*, i.e. that Planck scales for very distant objects (i.e. for early times) are so small that they may have no observational consequences. The following calculations show that at least one of the current models on quantum gravity can fit the observations reported in [23.24] if our hypothesis is taken into account.

According to equation (5) of [24], in order to observe the cited quantum effects in radiation of wavelength $\lambda$ coming from objects at a distance $L$, it has to be

$$1 \leq a_0 \frac{L}{\lambda}\left(\frac{l_p}{\lambda}\right)^{\alpha_e}$$

(29)

The parameters $a_0$ and $\alpha_e$ characterise the different quantum gravity theories (we have slightly changed the notation in order to avoid confusion of the exponent with the fine structure constant). Although other models have $\alpha_e=½$ (random-walk scenario), or $\alpha_e=2/3$ (holographic principle of Wheeler and Hawking; see [23]), the natural choice of $\alpha_e$ is 1 [32]. Then we have

$$a_0 L l_p \geq \lambda^2.$$

(30)

The coefficient $a_0$ is usually expected to be of order unity, but according to Amelino-Camelia [28], it can be a few orders of magnitude smaller. Substituting $l_p$ by $R/10^{61}$ (equation (4)), where $R$ refers to the moment when the light was emitted and making $L \cong R_o - R$, we obtain, for a representative $\lambda$ of $10^{-6}$m

$$a_0 \left(2 \times 10^{26} - R\right) R \geq 10^{49}.$$

(31)

This function of $R$ has a minimum at $R = 10^{26}$ m. Then a value of $a_0 = 10^{-3}$, compatible with [28], would yield no observable quantum effects at any scale. For instance, if we calculate the maximum value of $a_0$ compatible with the observations of the above cited SN 1994D, we get $a_0 = 0.15$. For lower values of $\alpha_e$ the constraints in $a_0$ are tighter and it is more difficult to reconcile the observations with the corresponding quantum gravity models. In our framework, certain models, such as those featuring $a_0 = 1$, appear to be incompatible with the above observations.

## VI. CONCLUSIONS

In this paper we have analysed the constancy of the product $\hbar c$ in the framework of *DeLightS*, a cosmological model with time-decreasing speed of light in which $c \propto R^{-1/2}$. This constancy of $\hbar c$ thus implies that $\hbar \propto R^{1/2}$. Furthermore, if $G$ is constant, it follows that $m_p$ is constant and that $l_p$ and $t_p$ change with time as a consequence of the variations of $c(t) \propto t^{-1/3}$ and $\hbar(t) \propto t^{1/3}$, in such a way that the value of the ratios in (4) stays constant along the history of the Universe. Thus *DeLightS* model, which solves the horizon problem without requiring inflation, and eliminates the need of interpreting the faintness of distant supernovae and radiogalaxies as an indication of an accelerated expansion driven by a mysterious dark energy (since it predicts for them a farther luminosity distance than constant-*c* models), also shows a formal appeal which is internally self-consistent and compatible with a range of observations.

Relation (4) is surprising by its simplicity, and uncovers a direct connection between the quantum scales and the cosmic scales. In thermodynamic terms, it implies the constancy of the entropy associated to the horizon of the observable Universe, as calculated from the Bekenstein-Hawking formula. This is consistent with the basic postulate of *DeLigthS* model: space-time framework is expanding at the same rate of the observable horizon, $c(t)$, so that no new information is gained or lost in the course of time. Perhaps this constancy of the Universe entropy could remove one of the classic arrows of time in physics, but another arrow of time appears, which is associated with the trends of temporal variations of *c* and $\hbar$.

Since $\hbar$ grows with the Universe, and quantum effects, such as particle-wave duality and the Heisenberg's principle of uncertainty, are proportional to Planck's constant, it is anticipated that such effects were smaller in the young Universe and will be more noticeable as the Universe grows old. In connection with this, we have seen how the *covariance hypothesis* helps to understand the absence of observable quantum structure in space-time because the farther away we observe looking for evidence of this structure, the younger the Universe was and, therefore, the smaller the quantum effects were.

On the other hand, $l_p$ and $t_p$ should be also changing with time, in such a way that they were smaller in the past. In this scenario, the Planck era, when cosmological models based on GR are thought to be invalid because quantum mechanics becomes imperative, is not reached until $t = R = 0$, i.e. not reached in fact. In this surprisingly simple way the problem associated with this theoretical barrier, needing the very difficult task –if possible- of developing a theory to unify GR and quantum mechanics, vanishes because quantum-gravity effects must disappear in the limit $l_p \to 0$ [27]. So, *the covariance hypothesis* avoids the confrontation of GR and quantum physics concerning the Planck era. If this idea is right, *DeLightS* model could be valid since the very beginning of time. Then the Big Bang could be infinitely hot, fast and dense, at least from a mathematical point of view.

Let's finally note that allowing for the variability of $c$ and $\hbar$, yields new dimensionless constants (14), (26)-(28) with remarkable consequences in the physical description of the Universe.

## ACKNOWLEDGMENTS

I am especially indebted to professor David Jou for helpful and encouraging discussions and for his open-minded support to these unconventional ideas.

---